\documentstyle[prl,multicol,aps,psfig]{revtex}

\begin{document}

\title{ Transition Temperature of the Homogeneous, Weakly Interacting 
Bose gas}
\author{Markus Holzmann$^{1,2}$ and Werner Krauth$^1$
\footnote{holzmann@physique.ens.fr;
krauth@physique.ens.fr, http://www.lps.ens.fr/$\tilde{\;}$krauth}}
\address{CNRS-Laboratoire de Physique Statistique$^{1}$ \\
and Laboratoire Kastler-Brossel$^{2}$
\footnote{Laboratoire Kastler-Brossel is a unit{\'{e}} de
recherche de l'Ecole Normale Sup{\'{e}}rieure et de l'Universit{\'{e}}
Pierre et Marie Curie, associ{\'{e}}e au CNRS.}\\
Ecole Normale Sup{\'{e}}rieure,
24, rue Lhomond, 75231 Paris Cedex 05, France}
\maketitle

\begin{abstract}
We present a Monte Carlo calculation for up to $N \sim 20\,000$ bosons
in $3$ D to determine the shift  of the transition temperature due
to small interactions $a$.  We generate independent configurations
of the  ideal gas. At finite $N$, the superfluid density changes
by a certain correlation function in the limit $a \rightarrow 0$;
the $N \rightarrow \infty$ limit is taken afterwards.  We argue
that our result is independent of the order of limits.  Detailed
knowledge of the non-interacting system for finite $N$ allows us
to avoid finite-size scaling assumptions.
\\ PACS numbers:
03.75.Fi, 02.70.Lq, 05.30.Jp \end{abstract}

\begin{multicols}{2}
\narrowtext
Feynman \cite{feynman} has provided us with a classic formula for
the partition function of the canonical noninteracting Bose gas.
It represents a ``path-integral without paths'', as they have been
integrated out. What remains is the memory of the cyclic structure
of the permutations that were needed to satisfy bosonic statistics:
\begin{equation}
Z_N =  \sum_{\{ m_k\}} {\cal P}( \{ m_k\}); \; \mbox{with}\; 
{\cal P}( \{ m_k\}) =  
\prod_{k=1}^{N} \frac{\rho_k^{m_k} }{m_k ! k^{m_k}}.    
\label{feynman} 
\end{equation}
The partitions $\{ m_k\}$ in eq.~(\ref{feynman})  decompose
permutations of the $N$ particles into exchange cycles ($m_i$ cycles
of length $i$ for all $1 \le i \le N $ with $\sum_k
k\;  m_k = N$). $\rho_k$ is a system-dependent weight
for cycles of length $k$.

In this paper we present an explicit Monte Carlo calculation for
up to $\sim 20\,000$ bosons in three dimensions, starting from 
eq.~(\ref{feynman}). The calculation allows us to determine unambiguously
the shift in the transition temperature $T_c$ for \textsl{weakly
interacting} bosons in the thermodynamic limit for an infinitesimal
s-wave scattering length $a$.  This fundamental question has lead
to quite a number of different and contradictory theoretical as well as
computational answers (cf, e.g., \cite{stoof,franck,ursell99}).

We will first use Eq.~(\ref{feynman}) and its generalizations to
determine very detailed properties of the finite-$N$ canonical Bose
gas in a box with periodic boundary conditions.  We then point out
that \textsl{all} information on the shift of $T_c$ for weakly
interacting gases is already contained in the noninteracting
system.  In the linear response regime (infinitesimal interaction),
it is a certain correlation function of the noninteracting system
which determines the shift in $T_c$.  This correlation is much too
complicated to be calculated directly, but we can  \textit{sample}
it, even for very large $N$.  To do so, we generate independent
bosonic configurations in the canonical ensemble. We have found a
solution (based on Feynman's formula Eq.~(\ref{feynman})) which
avoids Markov chain Monte Carlo methods.  In our two-step procedure, a partition
$\{m_k\}$ is generated with the correct probability ${\cal
P}(\{m_k\})$.  Then, a random boson configuration is constructed
for the given partition.

We stress that all our calculations are done very close to $T_c$,
so that the correlation length $\xi$ of any macroscopic sample is
much larger than the actual system size $L$ of the simulation. This
condition $L \ll \xi $  allows us to invoke the standard finite-size
scaling hypothesis\cite{zinn}, but also to take the $N \rightarrow
\infty$ limit after the limit $a \rightarrow 0$.

A key concept in the path integral representation of bosons is that
of a \textsl{winding number}. Consider first the density matrix
$\rho(r,r',\beta)$ of a single particle [$r=(x,y,z)$] at 
inverse temperature $\beta=1/T$. In a
three-dimensional cubic box of length $L$ with periodic boundary
conditions, $\rho(r,r',\beta) = \rho(x,x',\beta) \times 
\rho(y,y',\beta) \times  \rho(z,z',\beta)$ with, e.g.,
\begin{equation}
\rho(x,x',\beta)=
\sum_{w_x=- \infty}^{\infty}  
\frac{\exp[-(x-(x' + L w_x))^2/2 \beta]}{\sqrt{2 \pi \beta}}.
\label{winding} 
\end{equation}
In Eq.~(\ref{winding}), $x$ and $x'$ are to be taken within the
periodic box ($0<x,x'<L$).
\begin{figure}
\centerline{ \psfig{figure=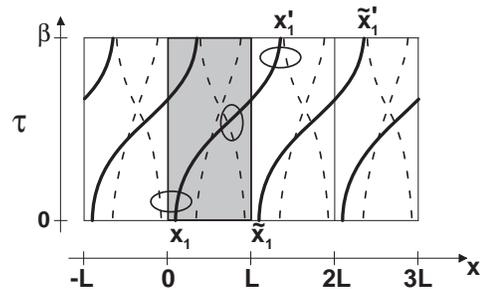,height=4.cm} }
\caption{ A one-dimensional periodic simulation box with three bosons.
Particles move in imaginary time $0< \tau <\beta$ and in periodic
space $0<x<L$. We use non-periodic coordinates. }
\end{figure}
It is more convenient  to adopt non-periodic coordinates ($-\infty
<x,x'< \infty$), as we will do from here on.  In Fig. 1, the path
drawn with a thick line can equivalently be tagged by $(x_1,x_1')$
or by  $(\tilde{x}_1,\tilde{x}_1')$.  This notation allows one to keep
track of the topology of paths without introducing intermediate
time steps $\tau$, even for very small systems.  With this convention,
the winding number of a configuration, $W=(W_x,W_y,W_z)$, is defined
as
\begin{equation}
W= \sum (r_i'-r_i)/L.
\label{winding2} 
\end{equation}
The winding number $W$ in Eq.~(\ref{winding2})  is the sum of the
(integer) winding numbers for each of the statistically uncorrelated
cycles which comprise the configuration.  The complete statistical
weight $\rho_k$  of a cycle of length $k$ [cf.  
Eq.~(\ref{feynman})] is given by  the sum of the weights $\rho_{k,w}$
for all winding numbers $w$:
\begin{equation}
\rho_k= \left[ \sum_{w=-\infty}^{\infty} \rho_{k,w} \right]^3; \; 
\rho_{k,w}=  \frac{L}{\sqrt{2 \pi k \beta }  }
\exp(-\frac{L^2 w^2}{2 k \beta}).
\label{winding_cycle} 
\end{equation}

Pollock and Ceperley \cite{ceperley} have obtained the result
\begin{equation}
\rho_s/\rho = \frac{<W^2> L^2}{3 \beta N },
\label{rhos}
\end{equation}
which connects the system's superfluid density $\rho_s/\rho$  to
the winding number in a rigorous fashion. 

It is possible to determine the mean square winding number $<W^2>$
from Eq.~(\ref{feynman}).  We first compute $<W^2>$
for a given partition $\{m_k\}$
\begin{equation}
<w^2>_{\{m_k\}} = \sum m_k <w^2>_k.
\label{separate}
\end{equation}
Here, $<w^2>_k$ is the mean with respect to cycles of length $k$,
$<w^2>_k = 3 [\sum_w \rho_{k,w} w^2 ] [\sum_w \rho_{k,w}  ]^2/
\rho_k $. This yields, by summation over partitions
\begin{equation}
<W^2> =  \sum \frac{\rho_k }{k} <w^2>_k Z_{N-k}/Z_N.
\label{wsqanal} 
\end{equation} 
We have also determined the probability distribution of $W_x$.

An analogous calculation  formally replaces $<w^2>_k \rightarrow k$
in Eq.~(\ref{separate}), which
becomes $\sum_k k\;  m_k = N$ [cf. Eq.~(\ref{feynman})].
Equation (\ref{wsqanal}) is transformed into
\begin{equation}
Z_N = \sum\rho_k Z_{N-k}/N.
\label{split-off}
\end{equation}
Equation (\ref{split-off}) allows the recursive calculation of the
partition function $Z_N$ if $Z_1, \ldots, Z_{N-1}$ are known
\cite{wilkens}. 

The same relation Eq.~(\ref{split-off}) allows us to identify
\begin{equation}
k<m_k>=\rho_k Z_{N-k}/Z_N
\label{cycle}
\end{equation}
as the mean number of particles in a 
cycle of length $k$.

 From a different point of view, the quantity $\sum_k k\,
m_k e^{-\beta\, k \,\epsilon_i} /\rho_k$ determines the occupation
number of single-particle energy levels $\epsilon_i$ for a given
partition $\{m_k\}$.  This  allows us to compute the average number
$<N_i>$ of particles occupying state $\epsilon_i$ in the bosonic
system:
\begin{equation}
<N_i>= \sum_{k=1}^N \left\{ e^{-\beta\, k \,\epsilon_i} \frac{Z_{N-k}}{Z_N}.
\right\}  
\label{occupation} 
\end{equation}
Eq. (\ref{occupation}) is of crucial importance: We find
that $N_0/N$, the condensate fraction, is different from the
superfluid fraction, as determined from Eqs~(\ref{rhos}) and
(\ref{wsqanal}) in a finite non-interacting system.

The term $\{ \}$ in Eq.~(\ref{occupation}) can be regarded as
the probability $P(n_i \ge k)$  of having at least $k$ particles
in state $\epsilon_i$.  Taking the sum over all states $i$, with
the use of Eq.~(\ref{cycle}), we can connect cycle statistics with
the usual occupation number representation:
\begin{equation}
k < m_k> = \sum_i P(n_i \ge k).
\end{equation} 
This curious result, which is of practical use in inhomogeneous
systems \cite{krauth96},  tells us that the discrete derivative of
the mean cycle numbers with respect to their length is given by 
the probability of having $k$ particles in the same single-particle energy
level.

Rescaled superfluid densities $N_i^{1/3} \rho_s/\rho$ (from 
Eqs~(\ref{wsqanal}) and (\ref{rhos})) are  plotted in Fig. 2a for
$N_1=37, N_2= 296,N_3= 2368, N_4=18\,944$ as a function of the rescaled
temperature $t=(T-T_c^{\infty})/T_c^{\infty}$, where $T_c^{\infty}$
is the critical temperature for $N \rightarrow \infty$.

\begin{figure}
\centerline{ \psfig{figure=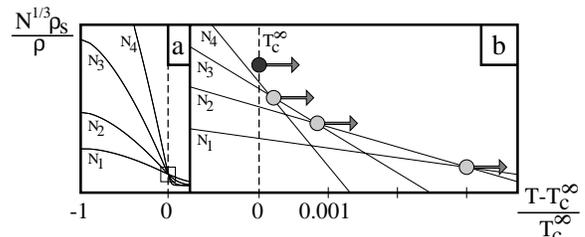,height=3cm} }
\caption{ Rescaled superfluid density of an ideal Bose gas.
The curves for different $N$ with $N_{i+1} = 8 N_i$ intersect approximately
at $T_c^{\infty}$, as shown in (a). The close-up view (b) 
reveals important differences. We determine
the shift of the intersection points as a function of the interaction
(light arrows). The dark arrow shows schematically  the extrapolated
shift in the thermodynamic limit.  }
\end{figure}
A finite-size scaling ansatz, which was used in previous Monte
Carlo work on the problem \cite{franck},  assumes that the curves
of $N_i^{1/3} \rho_s/\rho$ for a weakly interacting Bose gas should
intersect at the transition temperature, as they do approximately.
However, the small-scale Fig, 2b clearly shows the importance of
corrections to scaling (cf. \cite{pollock}) already for  the
noninteracting gas.  By continuity, the corrections to scaling
for the weakly interacting Bose gas must be important,
especially if the temperature shift due to interactions becomes small.

Our strategy greatly benefits from the solution Eq.~(\ref{wsqanal})
for the ideal gas.  We compute  the intersection point
($N_i^{1/3} \rho_s/\rho, t$) for two finite systems with $N_1$ and
$N_2= 8 N_1$ particles and determine how this point is shifted
under the influence of interactions (cf. Fig. 2b).  Our arbitrary
but fixed 
ratio $N_2/N_1=8$ facilitates the direct extrapolation
in $N_1$ to $N_1 \rightarrow \infty$.

To generate a random partition, we interpret the term $\rho_k Z_{N-k}/Z_N$ 
in  Eq.~(\ref{split-off})
as the probability
to \textsl{split off} a cycle of length $k$ from a configuration
of $N$ bosons, and to be left with a system of $N-k$ bosons.  We
can pick $k$ with probability $\sim  \rho_k Z_{N-k}$ with a simple
``tower of probabilities'' strategy \cite{Intro}.  Recursively, we
can thus generate an independent random partition $\{ m_k\}$ with
great speed.   The recursion stops as soon as we have split off a
cycle of length $j$ from a system with $j$ particles.

To go from a random partition to a random \textit{configuration}, we may
treat each cycle separately. For a cycle of length $k$, we select a
winding number $w_x$ with probability $\rho_{k,w_x}$ [cf. 
Eq.~(\ref{winding_cycle})], and analogously for $w_y$ and $w_z$.  Towers
of probabilities are again used. The cycle starts at a random
position $r=(x,y,z)$ with $0<x,y,z<L$, and ends at $r'=(x+w_x
L,y+w_y L,z+w_z L)$. Intermediate points are filled in with the
appropriate L\'{e}vy construction \cite{ceperley}.  We have
tested our  algorithm successfully against the known results (cf.
Fig. 2).

We thus generate  independent free boson path-integral configurations
by a method very different from what is usually done in path-integral
(Markov-chain) Quantum Monte Carlo calculations,
but with an equivalent outcome:
Any appropriate operator  is sampled with the probability
\begin{equation}
<{\cal O}>_0  =  \frac {\sum_P \int dR 
\left[\prod_{i} \rho(r_i,r_i')\right] {\cal O} }
{\sum_P \int dR \left[\prod_{i} \rho(r_i,r_i')\right]}.
\end{equation}
Here, $\sum_P$ indicates the summation over all permutations $P$, and
$r_i'$ is the position of the particle $P(i)$.
In the presence of interactions, the statistical weight of each
configuration is no longer given by the product of the one-particle
density matrices $\pi_0 = \left[\prod_i \rho(r_i,r_i')\right]$.
To lowest order in the interaction, the density matrix is exclusively
modified  by s-wave scattering. Likewise, only binary
collisions need to be kept.  This means that the correct statistical
weight is given by
\begin{equation}
\pi_a(r_1,\ldots,r_N; r_1',\ldots,r_N') =\pi_0 
\prod_{i<j}g_{ij}(r_i,r_j,r_i',r_j').
\label{2-prop}
\end{equation}
The contribution of collisions is to lowest order in $a$ 
\begin{equation}
\prod_{i<j}g_{ij} = 1 - a \sum_{i<j}c_{ij}
\end{equation}
with
$$
c_{ij} = \left( |r_{ij}|^{-1} +
|r_{ij}'|^{-1}\right)
\exp[|r_{ij}|\, |r_{ij}'|\, (1 + \cos \gamma_{ij} )/2\beta].
$$
Here, $r_{ij}=r_i-r_j$ and 
$\gamma_{ij}$ is the angle between $r_{ij}$ and $r_{ij}'$.
Equation (\ref{2-prop}) corresponds to the  popular Path-Integral 
Monte Carlo ``action'' with the following important modifications:
\textit{i)} no interior time-slices are needed, 
\textit{ii)} the interaction may be treated
on the s-wave level, and 
\textit{iii)} the interaction may be expanded in $a$. 
For a consistent evaluation of the interaction with periodic 
boundary conditions,
as schematically represented in Fig. 1, it is best to
sum over \textit{all} pairs $i<j$ shown, with the condition that
$r_i$ be in the original simulation box (shaded in gray). As
indicated by the small circles in Fig. 1, $c_{ij}$ may have
important contributions stemming from more than one representative
of the path $(r_j,r_j')$, especially for small systems.  Of course,
a cutoff procedure can be installed.

We now find for the  mean-square winding number in the interacting system
\begin{equation}
<W^2>_a = \frac
{<(1 - a C) W^2>_0}
{<( 1 - a C)>_0},
\end{equation} 
where we put $C= \sum_{i<j} c_{ij}$.
Expanding in $a$,  this yields
\begin{equation}
<W^2>_a - <W^2>_0  =
 - a <(\Delta W^2) (\Delta C)>_0,
\label{correlation}
\end{equation}
where $\Delta {\cal O}={\cal O} - <{\cal O}>_0$ \cite{footnoise}.

For a finite system of $N$
bosons,  the shift in the superfluid density 
$\delta \rho_s = \rho_s(a) -\rho_s(0)$ can thus be proven to
be linear in $a$ 
\begin{equation}
\frac{\delta \rho_s}{\rho} = - \frac{X_N}{\beta N^{1/3}} a \rho^{1/3},
\label{deltarho_s}
\end{equation}
with $X_N=<(\Delta W^2) (\Delta C)>_0/(3 \rho) $.

To determine quantitatively the shift of the intersection points
in Fig. 2, we expand the ideal gas superfluid density around
the intersection temperature $T_s$ of two systems with $N$ and $8 N$
bosons at the same density:
\begin{equation}
\rho_s/\rho(T) N^{1/3} = \rho_s/\rho(T_s) N^{1/3} + \alpha_N \times [T- T_s]. 
\label{linear}
\end{equation}
In this formula, the linear expansion coefficients can be
computed.  With interactions, only $\rho_s/\rho (T_s) N^{1/3}$ is modified
to linear order in $a$. 
$\alpha(N)$ remains unchanged, as we restrict the expansion to  
$|T-T_s|/T_s \sim a \rho^{1/3}$. 
We find the  new intersection point of the two systems  to be  
shifted in temperature as 
\begin{equation}
\frac{\Delta T_s}{T_s(0)} := \frac{T_s(a) - T_s(0)}{T_s(0)}
 = \frac{X_{8 N} - X_{N}}
{\alpha_{8 N} - \alpha_{N}} a \rho^{1/3}.
\label{shift}
\end{equation}
We have also computed the shift in $\rho_s/\rho$, 
but found out only that it  must be extremely small. We are unaware of
any fundamental reason for a vanishing shift in this quantity.
In Fig. 3, we plot the shift $\Delta T_s$ for different system sizes
ranging from  $(N_1,N_2)=(37,296)$ to $(2368,18\,944)$ vs $N_1^{-1/2}$. We have
not attempted a thorough
analysis of the finite-size effects, which already appear negligible 
for our largest systems. 

\begin{figure}
\centerline{ \psfig{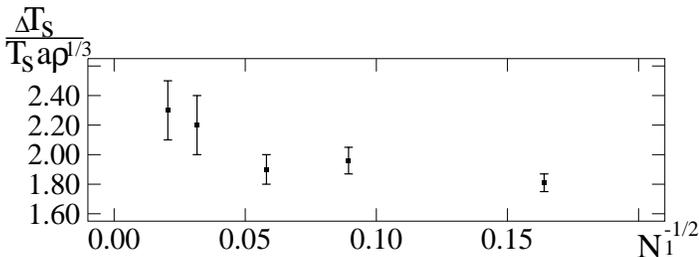} }
\caption{Shift of the intersection temperature
$\Delta T_s/[T_s(0)\; a \rho^{1/3}]$ as a function of $N_1^{-1/2}$
($N_2 = 8 N_1$). The system sizes are $(N_1,N_2)=(37,296)$,
$(125,1000)$, $(296,2368)$, $(1000,8000)$, and  $(2368,18\,944)$.}
\end{figure}

We conclude that the transition temperature of the weakly interacting
Bose gas increases linearly in the scattering length $a$ by an amount of
\begin{equation}
\frac{\Delta T_C}{T_C}=(2.3 \pm 0.25) a \rho^{1/3}.
\label{final} 
\end{equation}

Our result Eq.~(\ref{final}) is almost an order of magnitude larger than 
what was found in a previous Monte Carlo calculation
\cite{franck}. However, this calculation
was restricted to very small particle numbers and it used a problematic 
finite-size scaling ansatz, as pointed out. The agreement of
Eq.~(\ref{final}) with the 
renormalization group calculation \cite{stoof} seems to be quite good.

It is very interesting to understand whether the result 
Eq.~(\ref{final}) directly applies to the current Bose-Einstein
condensation experiments (cf, e.g. \cite{anderson95,davis95}).
In earlier papers \cite{krauth96,holzmann99}, we have pointed out the 
particularities of these finite systems in external potentials
(cf. \cite{dalfovo} for a general overview).  
Notwithstanding the differences between the two systems, a 
relevant parameter is for both cases  $a\rho^{1/3}$, where the 
maximum density (at the center of the trap)  at the transition point
must be taken in the 
inhomogeneous case. The experimental value  is of the order
$a\rho^{1/3} \sim 0.02 $. 

Within our method, we can also study finite values of the interaction,
even though we no longer compute a correlation function,
and also have to introduce interior time slices. Contributions 
beyond s-wave scattering need to be monitored, as we have in 
\cite{holzmann99}.
For $N=125$ bosons we have found  agreement with the linear reponse formula
Eq.~(\ref{correlation})
up to $a \rho^{1/3} \lesssim 0.005$, but a $15 \%$ decrease for the
full treatment for $a \rho^{1/3} = 0.023$.
A detailed investigation of this question  
goes beyond the scope of this paper.

In conclusion, it is worth noting that we have encountered none of the
difficulties which usually haunt boson calculations:  We work
in the canonical ensemble; therefore, the fluctuation anomaly of
the grand-canonical Bose gas plays no role. The density remains
automatically constant as a function of $a$ so that an expansion in
$a \rho^{1/3}$ is well-defined.  At finite $N$, we
can furthermore prove that the shift in $\rho_s/\rho$ is linear in
the interaction parameter. We have also consistently approached the
weakly interacting system from the vantage point of the ideal gas. This
allows us to obtain the crucial information on exactly where
to do our simulation (cf. Fig. 2).  Finally, our  extremely powerful
sampling algorithm has allowed us to partially dispel the
curse of Monte Carlo simulations: limitations to small system sizes.

We thank Boris V. Svistunov for discussions.
M. H. acknowledges support by the Deutscher Akade\-mi\-scher
Austauschdienst. The FORTRAN programs used in this work 
are made available (from MH or WK).


\begin{thebibliography}{10}
\bibitem{footnoise} 
As it stands, eq.~(\ref{correlation}) is extremely sensitive to
sampling noise both in $C$ and in $W^2$. 
The latter can be completely eliminated by an explicit  calculation 
of the probability distribution of $P(W_x)$, along the lines of
eq.~(\ref{wsqanal}).
\end{thebibliography}

\end{multicols}
\end{document}